\documentclass[letterpaper,twocolumn,10pt]{article}
\usepackage{usenix2024_SOUPS}

\usepackage{tikz}
\usepackage{amsmath}
\usepackage{subcaption}
\usepackage{multirow}

\begin{document}
%-------------------------------------------------------------------------------

%don't want date printed
\date{}

\title{\Large \bf Stop Abandoning Me: Exploring the Landscape of \\Unmaintained Intimate Partner Abuse Support Applications}

\def\plainauthor{Turk et al.}

\author{
{\rm Ivy Turk}\\
Ruhr-Universität Bochum
\and
{\rm  Jasmin Wyss}\\
Ruhr-Universität Bochum
\and
{\rm  Rebekah Overdorf}\\
Ruhr-Universität Bochum\\Research Center Trustworthy Data Science and\\Security in University Alliance Ruhr
} % end author

\maketitle
\thecopyright

%-------------------------------------------------------------------------------
\section{Introduction} 
%-------------------------------------------------------------------------------
% Framing
Those experiencing intimate partner abuse (IPA) by an abuser with physical access to them as well as their devices are hard to reach and difficult to protect. This challenge does not stop well-meaning people from developing creative and beneficial support tools for those who need them most~\cite{sumra23appdv}. Although the creation of these tools is often strongly motivated and well intentioned, there is significantly less consideration and support for these tools long term. This leads to a loss of support over time, broken functionality and outdated information, and an overall lack of maintenance~\cite{ahmed26brokenapps}. In essence, the created tools become \textit{abandoned}, to the severe detriment of vulnerable users who rely on their continued existence.

There are a multitude of reasons why an application may lose support over time. An app may simply be a prototype, never intended for widespread use, e.g.~\cite{sultana21unmochon,mangeard24warne,goyal22harassment}. A tool may be tied to fixed funding, and running costs cannot be paid after the funding period ends. The technical leads on a project may provide the tool to support organisations to use, but become busy with other projects and leave the organisation to manage the tool by themselves, which is not always possible~\cite{taylor13handover}. In many cases, the loss of technical support for a tool is no-one's fault. But this matters little for the users who lose access to beneficial systems.

% Impact
App abandonment has a severe impact on users. Survivors of abuse often use support tools for emergency contacts, support networks, and evidence collection. Establishing a stable connection to emergency support is incredibly difficult in partner abuse scenarios due to surveillance~\cite{tseng20ipsforums,turk25spyot,ceccio23spydevices} and shared access to devices~\cite{doerfler2024privacyvstransparency}, and losing an emergency support app isolates survivors. Similarly, gathering evidence of long-term abuse takes extensive effort~\cite{stephenson26sherloc}, which is undone in an instant if the evidence storage app goes down or the application is no longer compatible with a newer OS version. These harms need to be mitigated, and while these applications can help users, their abandonment leaves survivors back where they started, if not worse off~\cite{taylor13handover}.

In this ongoing research, we investigate the landscape of abandoned applications through a measurement study of publicly available and discussed support tools online. Our research question is:
\begin{itemize}
    \item What proportion of domestic abuse support apps have lost support or are no longer available?

\end{itemize}

%-------------------------------------------------------------------------------
\section{Methods}
%-------------------------------------------------------------------------------
We explored the landscape of support applications for people who experience abuse and whose abuser is presumed to have access to their devices through a four stage process. First, we define search terms for app stores, search engines, and publication databases to identify support applications. We then construct a dataset of these that are either available, publicly discussed in academic or journalistic outlets, or otherwise have left discoverable traces of their existence on the web. We annotate metadata and descriptive data about these tools which are then analysed for correlations between application characteristics and liveness.

\paragraph{Data Collection}
We perform a series of searches the following platforms hosting our target application types: Google Play (Android apps), Apple App Store (iOS apps), Google Scholar (academic outputs), GitHub (code repositories), and Google Search (web applications and tools fitting other categories). We used snowball sampling to identify further tools matching our criteria by looking at academic citations, suggested apps on the mobile app stores, and lists of tools found via web searches. The full list of search terms are provided in Appendix~\ref{apx:search-terms}.

\paragraph{Annotation}
We annotate our dataset with two types of information: system metadata, and descriptive labels. Our metadata includes publicly available information on the tools' publication dates, platforms distributing the tool, most recent updates and review dates, and mean review scores. Our descriptive labels categorise applications by target audience, type of support provided (described in detail in Appendix~\ref{apx:category-list}), and the liveness status of the tool. All annotations are objective, and therefore do not require multiple annotators to label. Our dataset was compiled by the lead author, and checked by other authors.

We classify applications as one of ``live'', ``No updates in \{1,2,3+\} years'', ``On request only'', ``Paid access only'', ``Not publicly distributed'', and ``Down''. We later group these labels for simplicity into ``Live'', ``Restricted Access'' (on request only / paid access only), ``Abandoned'' (no updates in 1+ years), and ``Unavailable'' (down / not distributed). According to application analytics company 42matters~\cite{appupdatesstats}, 96\% of the top 1\,000 Google Play apps and 97\% of the top 1\,000 iOS apps were updated in the last year. Therefore, we chose to use one year since the latest update as a threshold for ``abandoned'' applications, as we can expect live apps to be updated at least yearly.

\paragraph{Analysis}
Our annotated dataset consists solely of measurable data about the applications, and therefore we use different statistical measures to explore it. In future work, we plan on using a logistic regression model to look for correlations between types of application and their current liveness status. 
In our current research, this model is limited as many categories have too few matching applications, and therefore we instead leave this for future analysis once we have grown our dataset to a sufficient size.

\paragraph{Ethics}
Our dataset consists of no human data, and only publicly available information about applications which are intended to be public. Therefore, we determined that an ethical review was not necessary, as long as sensible precautions were implemented. To avoid undue strain on distribution servers and networks, we performed data collection manually, which limited the number of queries performed. Regarding the wellbeing of team members, we made sure to keep an open communication channel between co-authors to alleviate undue psychological strain.

%-------------------------------------------------------------------------------
\section{Preliminary Results}
%-------------------------------------------------------------------------------
Our results so far focus on statistics of the applications in our preliminary database. We describe the application dataset as of August 2026, the types of apps found according to our classifications, and the ``liveness'' of apps in the dataset. Our dataset is the result of search queries in English and Spanish, with future extensions planned for French and German. We used a VPN to obtain results from the UK, USA, Australia, Spain, Mexico, and Colombia.

\paragraph{Application Dataset}

Our dataset contains 197 support tools, including 135 Android apps, 111 iOS apps, and 8 web applications. Notably, 46 tools were exclusively available on one mobile operating system, and 13 were not distributed via official app stores. 

\paragraph{Types of Support Provided}
The majority of applications identified are \textit{information} apps (139 / 197, 70.6\%) which assist the user in understanding abusive behaviours, the support available, and connecting with a support network. We identify fewer evidence collection/distribution (39 / 197, 19.8\%) and emergency support (74 / 197, 37.6\%) applications that were at least partially targeted to people living through IPA. The detailed classifications of applications are provided in Table~\ref{tab:cats}\footnote{The ``any'' subcategory totals the unique applications with at least one subcategory within the overarching category}.

\begin{table}[t]
    \centering
    \caption{Categories of Applications}
    \begin{tabular}{|c|c|c|} \hline
        Category                     & Sub-Category & Count \\ \hline
        \multirow{6}{*}{Information} & Education    & 81        \\
                                     & Services     & 54    \\
                                     & Legal Advice &  7    \\
                                     & Networking   & 29    \\
                                     & Planning     & 30    \\ \cline{2-3}
                                     & Any          & 139    \\ \hline
        \multirow{3}{*}{Evidence}    & Collection   & 32    \\
                                     & Sharing      & 19    \\ \cline{2-3}
                                     & Any          & 39    \\ \hline
        \multirow{4}{*}{Emergency}   & Police       & 56    \\
                                     & Friends      & 23    \\
                                     & Sharing      & 18    \\ \cline{2-3}
                                     & Any          & 74    \\ \hline
    \end{tabular}
    \label{tab:cats}
\end{table}

\paragraph{Liveness}
\begin{figure*}
    \centering
    \begin{subfigure}[t]{0.5\textwidth}
        \centering
        \includegraphics[height=2in]{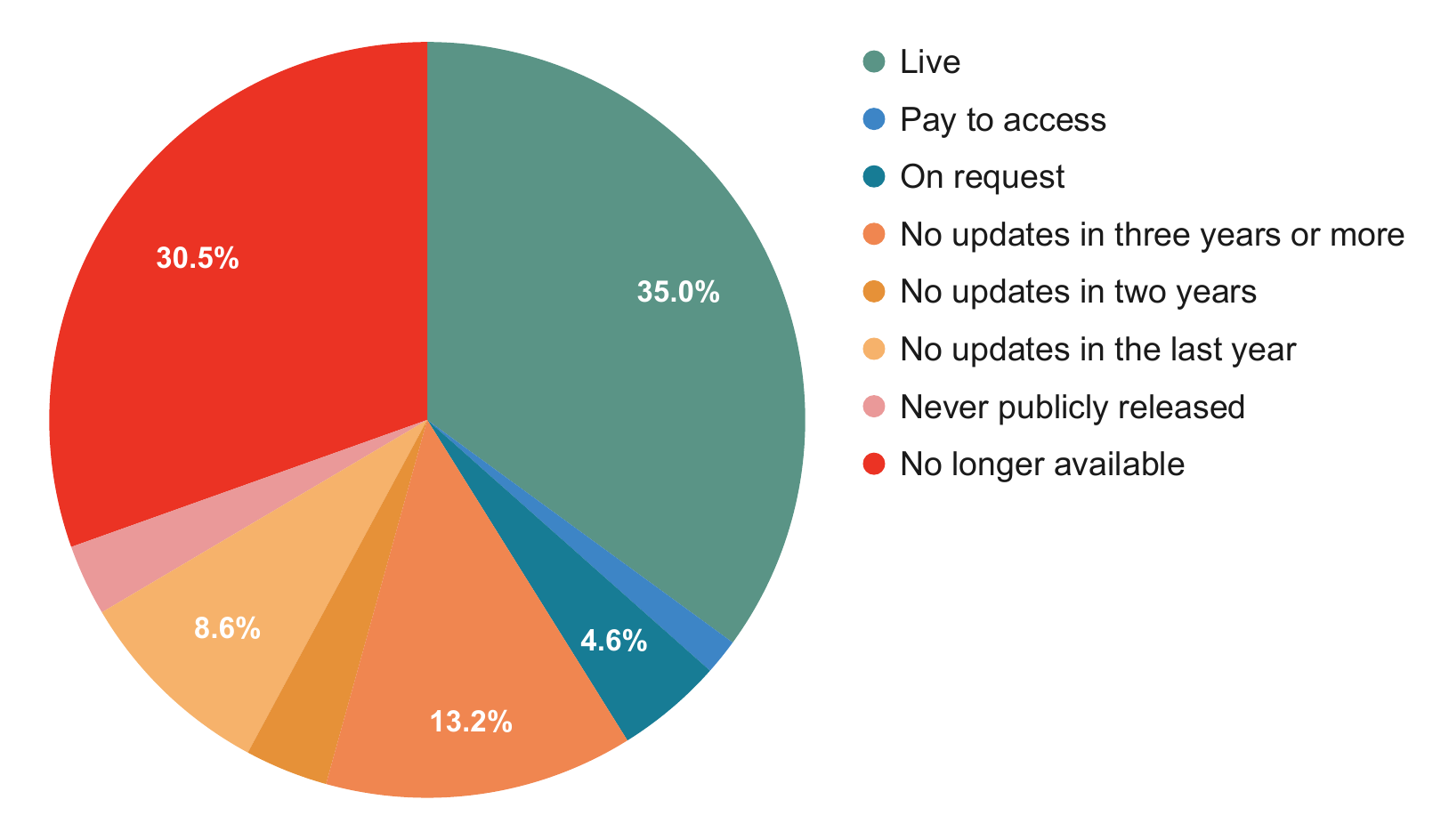}
        \caption{Availability of Applications}
    \end{subfigure}
    ~
    \begin{subfigure}[t]{0.4\textwidth}
        \centering
        \includegraphics[height=2in]{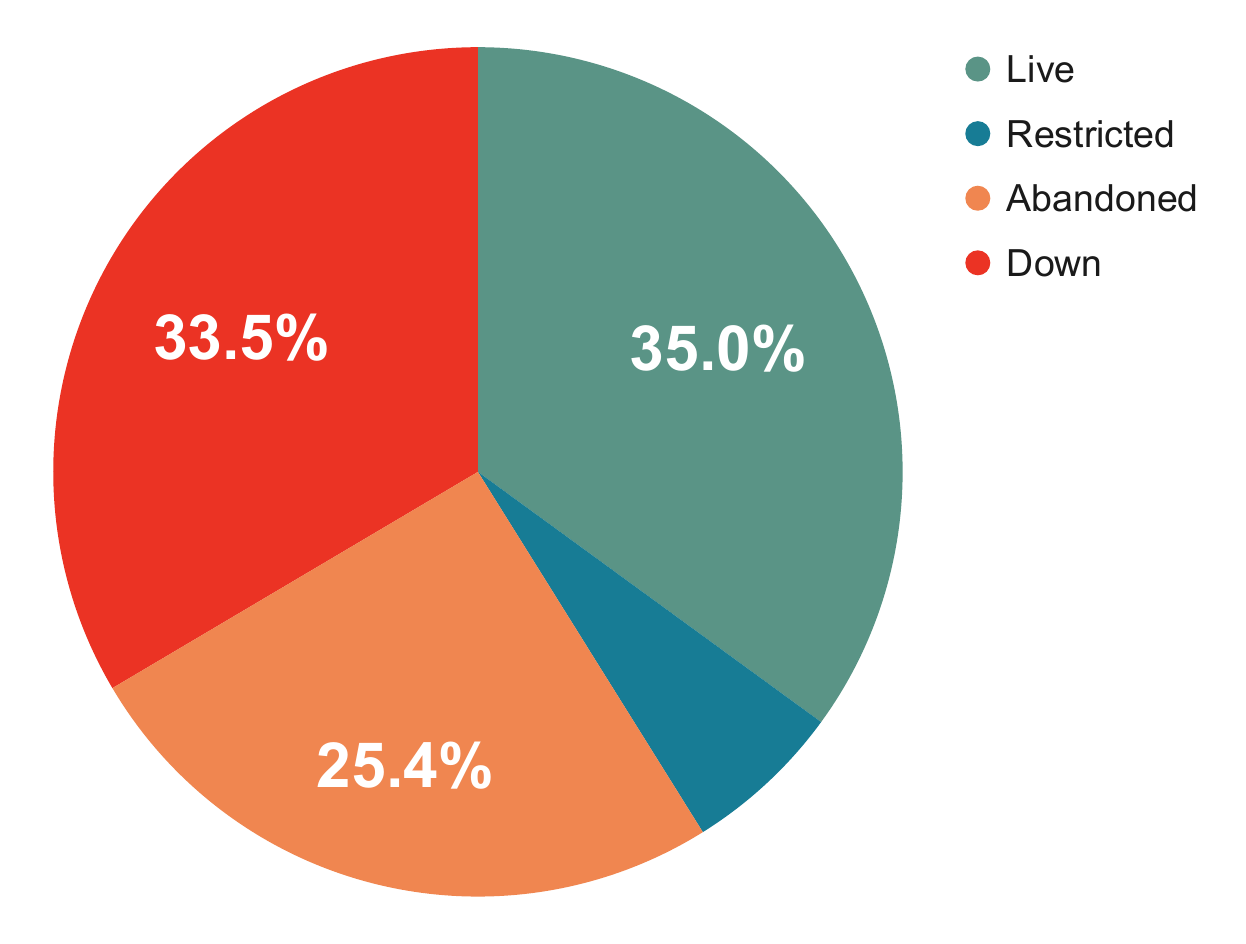}
        \caption{Categorised Availabilities}
    \end{subfigure}
    \caption{Current Availability of Support Applications in Dataset}
    \label{fig:placeholder}
\end{figure*}

Of the 197 applications in our dataset, 69 (35.0\%) were currently available and up to date, with a further 12 (6.1\%) available on request or with a paid subscription. Another 50 (25.4\%) were classified as abandoned, and 66 (33.5\%) are not available. Overall, \textbf{116 / 197 (58.9\%)} of applications in our dataset have lost support over time. This is a disconcertingly large figure, implying that half of all applications that someone uses for support will stop working over time.

\section{Follow-up Research}
We will continue with this research through extended data collection, aiming to include applications in languages other than English and Spanish. We will repeat our collection process with queries translated into French, German, and the spoken languages of additional recruited researchers. Furthermore, we will endeavour to maintain the database, collecting longitudinal "liveness data".

Assuming this provides sufficient data to enable statistical validity, we will use logistic regressions to identify the types of applications which are more likely to be abandoned. Additionally, we plan on exploring the impact of economic incentives on the long-term maintenance of the application using this data. 

We will also explore regional differences in application availability, distribution methods, and organisation affiliation as possible indicators of (an absence of) long-term support for these tools. We want to explore what kind of evidence-based policy and regulations recommendations we can issue based on our findings.

Going beyond the measurement of the available applications, we also plan on extending the project to evaluating the available applications. More specifically, we aim to measure the extent to which apps provide survivors the promised support functionalities.

%-------------------------------------------------------------------------------
\section*{Acknowledgments} 
%-------------------------------------------------------------------------------

The authors would like to thank Nimra Ahmed and Sophie Stephenson for inspiring this project through discussions at CHI 2026 in addition to their own research highlighting this issue. The authors thank the SOUPS 2026 attendees for their suggestions and feedback. Funded by the Deutsche Forschungsgemeinschaft (DFG, German Research Foundation) under Germany’s Excellence Strategy - EXC 2092 CASA - 390781972.

%-------------------------------------------------------------------------------
\bibliographystyle{plain}
\bibliography{abandoned_apps}

%-------------------------------------------------------------------------------
\appendix
\section{Application Search Terms}\label{apx:search-terms}
%-------------------------------------------------------------------------------
All searches were performed both with and without ``app'' as a suffix, for example ``Harassment support app'' and ``Harassment support''. These are translated into the authors' native and second languages.
\begin{itemize}
    \item Partner Abuse Terms
    \begin{itemize}
        \item Intimate partner abuse
        \item Intimate partner violence
        \item Domestic abuse
        \item Domestic violence
        \item Abuse support
        \item Violence support
        \item Abuse help
    \end{itemize}
    \item Forms of Abuse Terms
    \begin{itemize}
        \item Harassment support
        \item Surveillance support
        \item Spying support
        \item Spyware detection
        \item Spyware removal
    \end{itemize}
    \item Types of Support Terms
    \begin{itemize}
        \item Abuse information
        \item Abuse evidence
        \item Evidence Collection
        \item Abuse recording
        \item Abuse emergency 
        \item Emergency support
        \item Safety 
        \item Personal safety 
    \end{itemize}
\end{itemize}

\section{Application Support Types}
\subsection{Derivation of Categories}\label{apx:category-list}
We classify applications according to the form of support they provide. We build on the classifications defined by Sumra et al.~\cite{sumra23appdv}: legal information, self-assessment, informative, avoidance, and emergency assistance apps. The first four of these categories can be grouped under ``information'', while emergency assistance can be divided into Police / emergency services, contacting friends, and sharing information (such as current location) with contacts in both categories. We add an additional sub-category for information describing applications that aid in building a support network. There are also a wealth of evidence applications~\cite{stephenson26sherloc}, which we divide into collection and sharing.

%%%%%%%%%%%%%%%%%%%%%%%%%%%%%%%%%%%%%%%%%%%%%%%%%%%%%%%%%%%%%%%%%%%%%%%%%%%%%%%%
\end{document}